\documentclass[12pt]{article}
\usepackage[utf8]{inputenc}
\usepackage{setspace} 
\usepackage{enumitem} 
\usepackage[colorlinks=true,urlcolor=blue]{hyperref}
\usepackage{graphicx}
\graphicspath{{figures/}}

\begin{document}

\newcommand{\speaker}[1]{{\em #1\\ }}
\newcommand{\eprint}[1]{\href{http://arxiv.org/abs/#1}{arXiv:#1}}
\renewcommand{\refname}{ \small References}
\let\OLDthebibliography\thebibliography
\renewcommand\thebibliography[1]{
  \vspace{-2ex}
  \OLDthebibliography{#1}
  \setlength{\parskip}{0pt}
  \setlength{\itemsep}{0pt plus 0.3ex}
  \vspace{-2ex}
}

\hyphenation{brems-strah-lung}

\thispagestyle{empty}
\begin{center}
{\LARGE
Summary Report\\
of the 721th WE-Heraeus-Seminar: \newline
Light Dark Matter Searches\\[8mm]
}

\begin{figure}[h]
\centering
\includegraphics[width=0.7\textwidth]{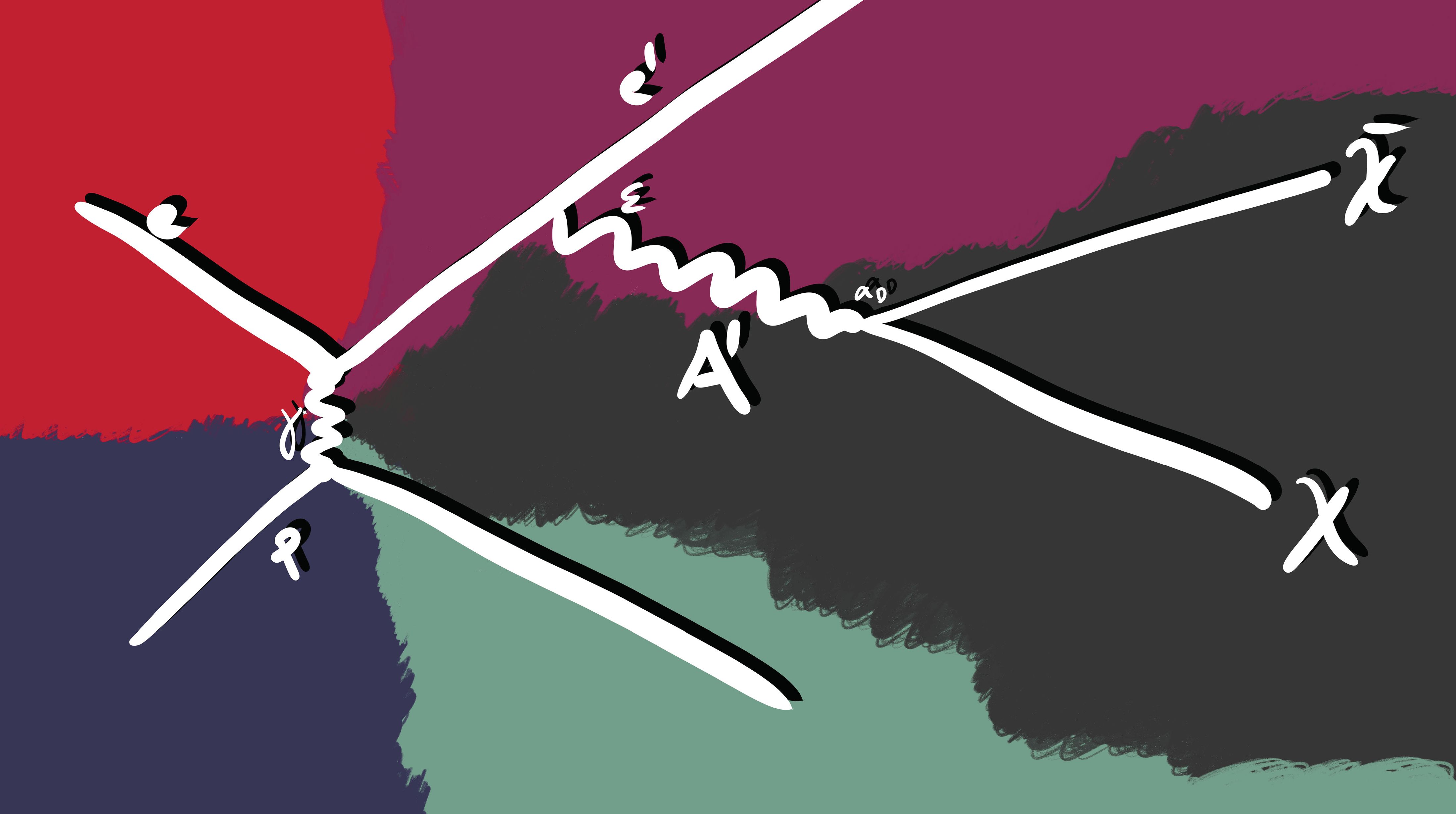}
\end{figure}

{\large Scientific Organizers:\\
{\em Patrick~Achenbach, Luca~Doria, and Marco~Battaglieri}\\[8mm]

Invited Speakers:\\ {\em Pedro~Schwaller, Claudia~Frugiuele, Felix~Kahlhoefer, Andrea~Celentano, 
Mirco~Christmann, Sören~Schlimme, Babette~Döbrich, Paolo~Valente, Ruth~Pöttgen, Maurik~Holtrop,
Heiko~Lacker, Stepan~Stepanyan, Tina~Pollmann, Federica~Petricca, Belina~von~Krosigk, Michelle~Galloway,
Torben~Ferber, Caterina~Doglioni, Dayong~Wang, Viktor~Zacek, Jan~C.~Bernauer, and Michael~Kohl}\\[8mm]

2021
}
\end{center}

\section{Aims and Scope}

From the 8 -- 11 June 2021 the 721th WE-Heraeus-Seminar was held online to discuss light dark matter (LDM) searches with national and international experts from experiments as well as from theory. The topic of this workshop is related to one of the biggest unanswered questions in physics: What is dark matter (DM)? Many observations suggest that the universe is filled with as yet unknown elementary particles that have in total about five times the mass of all ordinary matter. The long-standing search for such particles is focused at masses significantly above the proton mass and has up to now been unsuccessful. Consequently, there has been an intensified search for lighter particles that could be part of a dark sector (DS) of particle physics. Just as there are both matter particles and mediator particles of different forces in the SM, so far undiscovered matter particles in the DS could interact with each other through new forces. The DS mass scale could be comparable to the proton mass or below.

LDM would be very difficult to detect with high-energy colliders or with direct detection experiments using established techniques, so that accelerator-based DM searches with smaller, but dedicated experiments are becoming more important. In the seminar, many different ideas and methods of experimental verification were discussed, also because possible particles of a DS are energetically accessible at a number of accelerators worldwide. The research field includes, for example, measurements at the accelerator facilities MAMI in Mainz/Germany, DA$\Phi$NE in Frascati/Italy, Thomas Jefferson Lab in Virginia/US, J-PARC in Tokai/Japan, searches at the electron-positron experiments Belle~II at KEK/Japan, BaBar at SLAC/US, BESIII in Beijing/China and, last but not least, several ongoing and future projects at CERN. Some proposals are yet to be implemented, including those at the future accelerator MESA in Mainz. The capabilities of high-intensity electron and proton beams (also for neutrino physics experiments) enable unique opportunities for probing the DS. These accelerator-based approaches are complemented by new technological developments to detect the predicted cloud of DM particles in our Milky Way by collisions with sensitive detectors in underground laboratories such as Gran Sasso, deep beneath the Apennine Mountains of Italy. These experimental approaches are mostly complementary to searches for DM at the high-energy frontier at CERN.

One focus of the seminar was devoted to dark photons, the hypothetical counterparts of the quanta of the known electromagnetic interaction. However, this dark radiation is only one possible portal through which DS particles could interact with the SM. While most experiments provide exclusion limits for regions in mass and coupling strength, experimental evidence for a particle with mass 17 MeV, called X17, has been very controversial.

Although this was a virtual seminar, the online platform emulated a real scientific conference as closely as possible. It allowed not only for plenary sessions but also for poster sessions, and particularly encouraged personal interactions during the seminar. The numerous posters were produced with high quality by the young scientists and could be presented conveniently on the MeetAnyway platform. Because of the online format that became necessary and the widely separated time zones of the 22 invited speakers, the seminar had to be concentrated into a core time in the afternoon.

Pepe Gülker, Jennifer Geimer, and Goran Stanić were each awarded with a poster prize and an equal share of the prize money to highlight their outstanding presentations at the seminar.

We thank the Wilhelm und Else Heraeus-Stiftung for the organizational and financial support of the seminar and in particular Ms.~M.~Peklaj, whose  attentive and efficient support before, during, and after the conference was one of the key to success of the venue.
The Wilhelm und Else Heraeus-Stiftung is a private foundation that supports
research and education in science with an emphasis on physics. It is recognized as
Germany’s most important private institution funding physics. Some of the
activities of the foundation are carried out in close cooperation with the German
Physical Society (Deutsche Physikalische Gesellschaft). For detailed information
see \url{https://www.we-heraeus-stiftung.de}.
We are also thankful to Pepe~Gülker for the artwork on the cover page. 

In the following, the submitted abstracts of the invited speakers and the poster presenters are made available. The lectures are sorted into the five different session topics of the seminar, and have been complemented by points that were raised during the discussions. \\[8mm]

{\em Patrick~Achenbach, Luca~Doria, and Marco~Battaglieri}

\newpage
\section{Theory}

\subsection{\texorpdfstring{Light but Substantial:\\ DM Models Below the
WIMP Scale}{Light but Substantial: DM Models Below the
WIMP Scale}}

\speaker{Pedro Schwaller, Johannes Gutenberg University Mainz, Germany}

What is ``light” DM? An overview of the allowed mass range for DM was given, and then a few of the theoretically well motivated models were introduced that
predict DM masses significantly below the electroweak scale, such as axions,
sterile neutrinos, and asymmetric DM. Possible connections of
these models with the puzzle of the baryon asymmetry of the universe were highlighted, and then
a few recent ideas for probing such DM scenarios were presented, from fixed target
experiments to gravitational waves.

\subsection{Searching for Physics Beyond the SM
@ the Next Generation Neutrino Fixed Target
Experiments}

\speaker{Claudia Frugiuele, CERN, Switzerland}

Next generation neutrino oscillation experiments are multi-purpose observatories,
with a rich physics program beyond oscillation measurements. A special role is
played by their near detector facilities, which are particularly well-suited to search for
weakly coupled DS particles produced in the primary target. The
sensitivity for such scenarios of present and future facilities such as the SBN program
and DUNE were discussed.

\subsection{Self-interacting DM}

\speaker{Felix Kahlhoefer,
Institute for Theoretical Particle Physics and Cosmology,
RWTH Aachen University, Aachen, Germany}

Self-interactions between DM particles can affect the evolution of DM
halos, giving rise to observable effects in a wide range of astrophysical systems and
potentially explaining the puzzling observation of constant-density cores in dwarf
galaxies. To study these effects one needs to find a way to map the scattering of
elementary particles onto macroscopic scales. 

In this talk it was discussed how to
calculate transfer cross sections for the case that DM particles interact via the
exchange of a light mediator such as a dark photon with a particular focus on the
semi-classical regime. It was then shown how such scattering may give rise to an
effective drag force that can be implemented in numerical simulations of structure
formation in order to study merging galaxy clusters. Finally,  analytical
methods that can be used to describe the evolution of the central density of a DM halo and address the question whether DM self-interactions resolve
the small-scale problems of collision-less cold DM were discussed.

\subsection{Discussion of Theory}

\speaker{These points were compiled by the scientific organizers. Any statements, opinions, or conclusions contained herein do not necessarily represent the view of the speakers or participants. They are not meant to be complete or representative, but highlight the lively interchange of ideas in the seminar.}

\begin{itemize}[leftmargin=*]

\item Probably the best case scenario to probe a QCD-like DS at electron machines such as MESA would be a dark photon that couples this QCD-like DS to the SM. An alternative to the dark photon could be a heavy leptophilic mediator.

\item DM is very efficiently searched for in the NEAR detectors of neutrino experiments such as DUNE, where off-axis and on-axis options can be explored, depending on the experimental setup. DUNE is, for example, considering a moving near detector, although theoretical calculations show that the scope of such an arrangement is limited in terms of DM search reach. 
This scenario was first discussed in \eprint{1807.10842}, see Fig.~2. The counter-intuitive shape of the bound is a result of the specific model assumptions. If the mass of the light scalar is produced via the Higgs mechanism, there is an irreducible contribution to the decay $h \rightarrow \phi \phi$ that depends only on the scalar mass and not on the mixing angle. This leads to a vertical exclusion line, i.e.\ an upper bound on the scalar mass.

\item The explanation of relations such as the Tully-Fisher correlation is discussed in \eprint{1808.05695}.  

\item Cosmology (in particular the CMB spectrum) is the strongest supporter of particle DM. Modified gravity cannot reproduce the CMB spectrum. At smaller scales (galactic, ...) the two theories can still compete.

\item The possibility that some fraction of DM is in the form of black holes is very difficult to exclude. The possibility that {\em all} of DM is black holes can be tested through a range of astrophysical observations, depending very much on the specific mass range, but large chunks of this are presently still viable.
The best way to exclude black holes as the dominant form of DM might be to discover DM particles in the laboratory.

\end{itemize}

\section{Fixed-target Experiments}

\subsection{LDM Search with the BDX Experiment at Jefferson Laboratory}

\speaker{Andrea Celentano for the BDX Collaboration,
INFN-Genova, Genova, Italy}

The Beam Dump eXperiment (BDX) at Jefferson Laboratory is an electron-beam
thick-target experiment aimed to investigate the existence of LDM
particles in the MeV--GeV mas. The experiment will detect LDM particles produced by
the interaction of the primary CEBAF electron beam impinging on the Hall-A beam
dump with a downstream electromagnetic calorimeter, installed in a new dedicated
experimental hall. The expected signal signature is the electromagnetic shower
induced by the scattering of LDM particles with atomic electrons of the detector
material, resulting to a visible energy deposition. The electromagnetic calorimeter is
surrounded by a dual-layer active veto counter to reject cosmic backgrounds. Beam-related
backgrounds, on the other hand, are suppressed by passive shielding (iron
blocks) installed between the dump and the detector.

A proof-of-concept measurement has been performed in 2020 at JLAB in the present
unshielded configuration, with a 2.2\,GeV primary electron beam. The prototype
detector (BDX-MINI) consists of a PbWO$_4$ electromagnetic calorimeter, surrounded
by a layer of tungsten shielding and two hermetic plastic scintillator veto systems.
The sensitivity of the test will reach some of the best limits to date for selected
regions of the LDM parameters space.

In this talk, after the BDX physics case was presented, the experiment design
and foreseen performances were discussed. Finally, some preliminary results from the
BDX-MINI measurement campaign were presented.

\subsection{The DarkMESA Experiment}

\speaker{Mirco Christmann for the MAGIX Collaboration, Institute for Nuclear Physics and
Helmholtz Institute Mainz, Johannes Gutenberg University Mainz, Germany}

At the Institute for Nuclear Physics in Mainz the new electron accelerator MESA will go
into operation within the next years. In the extracted beam operation (150\,MeV, 150\,$\mu$A)
the P2 experiment will measure the weak mixing angle in electron-proton scattering in
10,000 hours operation time. Therefore, the high-power beam dump of this experiment
is ideally suited for a parasitic DS experiment --- DarkMESA [1,2].

The experiment is designed for the detection of LDM which in
the simplest model couples to a massive vector particle, the dark photon 
$\gamma^\prime$. It can
potentially be produced in the P2 beam dump by a process analogous to photon
bremsstrahlung and can then decay in DM particle pairs $\chi\bar{\chi}$. A fraction
of them scatter off electrons or nuclei in the DarkMESA calorimeter [3].

In 2018, possible calorimeter materials were tested at MAMI  with electrons below 14
MeV. During this beam tests PbF$_2$ and the lead glass of type SF5 from Schott performed best. This result
was consistent with a {\tt Geant4} optical photon study [4,5].

For DarkMESA, about 1,000 PbF$_2$ crystals from the previous A4 experiment
at MAMI, with a total active volume of 0.13\,m$^3$, will be used. In a further stage the active volume will
be increased by adding more than 1,000 Pb-glass blocks (0.58\,m$^3$) from the previous
WA98 experiment at CERN.

Within a {\tt MadGraph} and {\tt Geant4} simulation the accessible parameter space was estimated.
The experimental setup was optimized and further concepts were investigated.
DarkMESA-Drift is such an additional approach. A directional Time Projection Chamber
(TPC) filled with CS$_2$ at low pressure serves as DM detector. With the nuclear recoil
threshold being in the keV range, the accessible parameter space can be extended [2].

Simulation studies and experimental studies on beam-related and beam-unrelated
backgrounds at DarkMESA were presented and the current status of a prototype detector
array including a veto system as well as a veto concept for the final DarkMESA
experiment was discussed.

\subsection{Dark Photon Searches at MAGIX}

\speaker{Sören Schlimme for the MAGIX-Collaboration,
Institute for Nuclear Physics, Johannes Gutenberg University Mainz, Germany}

From the very beginning, the research topic of the dark photon was an important motivation
for the construction of the Mainz Energy-Recovering Superconducting Accelerator
MESA [1], an electron accelerator which will come into operation at the Institute
for Nuclear Physics of the Mainz University in a few years.

Dark photon searches will be performed at the MAinz Gas Internal target eXperiment
MAGIX [2], which will be installed in the Energy Recovery Linac (ERL) arc of MESA.
The ERL mode provides a power-efficient beam acceleration with a maximum beam
energy of 105\,MeV and a maximum beam current of 1000\,$\mu$A or higher.

The internal, windowless MAGIX gas jet target [3] can be operated with different target
gases such as hydrogen or xenon. The setup comprises two high-resolution magnetic
spectrometers which will be used for the detection of scattered electrons and produced
particles. Their focal planes will be equipped with TPCs (time projection chambers)
with GEM (gas electron multiplier) readout for tracking, and with scintillation detectors
for trigger, timing, and particle identification purposes. Additional detectors dedicated
to the measurement of low-energetic recoil nuclei complement the spectrometers and
will be mounted in the target chamber.

With the MAGIX setup, the searches for dark photons, which assume a production of
the dark photon through a mechanism similar to the bremsstrahlung process, will be
extended to lower dark photon masses than were probed before in Mainz at the A1
spectrometer facility [4,5]. In this type of experiments a dark photon 
$\gamma^\prime$ could radiatively
be produced off a nuclear target $Z$ via the reaction $e^-Z \rightarrow e^-Z\gamma^\prime$. If the dark photon decays
into SM particles (visible decay), e.g.\ 
$\gamma^\prime \rightarrow e^+e^-$ the electron/positron
final state can be detected in coincidence in the two spectrometers. A peak-search on
the QED background can thus be performed.

These searches will be extended to invisible decay channels, in which a dark photon
decays by 
$\gamma^\prime \rightarrow \chi\bar{\chi}$ in a pair of possible LDM particles, requiring a missing
mass analysis of the scattered electron in coincidence with the recoil nucleus.

\subsection{Exotics Searches at NA62}

\speaker{Babette Döbrich, CERN, Switzerland}

Thanks to its high intensity beam and detector performance (redundant
particle identification capability, extremely efficient veto system and high resolution measurements of momentum, time, and energy), NA62~[1] can achieve
sensitivities to long-lived light mediators in a variety of new-physics scenarios.
This talk covered some phenomenological~[2] and technical~[3] aspects of exotics searches at NA62 highlighting the need of a close theory/experiment interface to optimize search
strategies.

\subsection{Searching Light Dark Particles in Positron Annihilations at PADME}

\speaker{Paolo Valente for the PADME Collaboration,
Istituto Nazionale di Fisica Nucleare, Sezione di Roma, Rome, Italy}

The lack of direct experimental observation of DM candidates at the electroweak scale so
far could be justified with the introduction of a DS that can only feebly interact with the
ordinary matter. A simple possibility would be an additional $U(1)$ symmetry, which corresponding
gauge boson is a massive vector particle, a dark photon.

Several experiments are searching for a dark photon in the visible (to lepton pairs) or invisible
decays (to lighter, DS particles).
PADME is the first experiment using the annihilation of a positron beam against a thin target, as a
production channel for the dark photon, and the missing mass technique to discover the dark
photon as a peak above a smooth background.

In 2019--2020 PADME has collected a sample of $5\cdot 10^{12}$ positron on target annihilations, in order to
probe a mass range up to 21\,MeV$/c^2$. Thanks to the excellent performance of the photon detectors,
also other processes can be studied, like the visible decays of a axion-like particle, or the
hypothetical X17 particle advocated for interpreting the anomalies in $^8$Be and $^4$He transitions.

\subsection{The LDM eXperiment -- Status, Plans
\& Prospects}

\speaker{Ruth Pöttgen, Lund University, Lund, Sweden}

An elegant explanation for the origin and observed abundance of DM in the
Universe is the thermal freeze-out mechanism. Within this mechanism, possible
masses for DM particle candidates are restricted approximately to the MeV--TeV range. The GeV--TeV mass range is being explored intensely by a variety of
experiments searching for Weakly Interacting Massive Particles. The sub-GeV region
occurs naturally in hidden sector DM models, but has been tested much less
by experiments to date. Exploring this mass range is imperative as part of a
comprehensive DM search programme, but requires new experimental
approaches.

The freeze-out mechanism assumes a non-gravitational interaction between dark
and ordinary matter, which necessarily implies a production mechanism for DM at accelerator experiments. Recent advancements in particle accelerators and
detectors in combination with software developments like machine learning
techniques open new possibilities to observe such processes.

The planned LDM eXperiment [1] (LDMX) is an electron-beam, fixed-target
experiment that exploits these developments, enabling us to observe
processes orders of magnitudes rarer than what is detectable today. The key to this
is a multi-GeV beam providing a few electrons 46-million times per second, and a
detector that monitors how each individual electron interacts in the target — for up to
10$^{16}$ electrons. First beam for commissioning the experiment is expected in 2023 at
SLAC, Stanford, marking the starting point of a first data-taking period of about 1.5
years. A second run with higher beam-energy and -intensity is foreseen soon
thereafter, either at SLAC or potentially CERN.

This presentation gave an overview of the different components of the LDMX
detector concept, the main experimental challenges and how they are addressed. It also briefly described the special requirements on the beam needed and how
these are met. Finally, it discussed projected sensitivities and possible future
upgrades towards covering a large portion of the viable phase-space for sub-GeV
thermal relic DM and other models.

\subsection{The Heavy Photon Search Experiment at JLab}

\speaker{Maurik Holtrop for the HPS Collaboration, University of New Hampshire, Durham, USA}

The Heavy Photon Search (HPS) experiment at Jefferson Lab is searching
for a new $U(1)$ vector boson (``heavy photon”, ``dark photon” or $A^\prime$) in the
mass range of 20--500 MeV$/c^2$. An $A^\prime$ in this mass region is natural in hidden
sector models of light, thermal DM. The $A^\prime$ couples to the ordinary
photon through kinetic mixing, which induces its coupling to electric charge.
Since heavy photons couple to electrons, they can be produced through a process
analogous to bremsstrahlung, subsequently decaying to an $e^+e^-$ pair,
which can be observed as a narrow resonance above the dominant QED trident
background. For suitably small couplings, heavy photons travel detectable
distances before decaying, providing a second signature. HPS accesses unexplored
regions in the mass-coupling parameter space.

The experiment uses the CEBAF electron beam located at Jefferson Lab to
accelerate electrons which are then incident on a thin tungsten target. The outgoing
$e^+e^-$ pair is detected in a compact, large acceptance forward spectrometer
consisting of a silicon vertex tracker, a hodoscope, and a lead tungstate
electromagnetic calorimeter.

HPS conducted successful engineering runs in the spring of 2015 using a
1.056\,GeV, 50\,nA beam and in the spring of 2016 using a 2.3\,GeV, 200\,nA
beam, and an extended physics run using a 4.5\,GeV beam during the summer
of 2019. 

This talk presented the results of the 2015 run, preliminary results of
the 2016 and 2019 runs, and prospects for future runs.

\subsection{LDM Search with SHiP}

\speaker{Heiko Lacker, Institute of Physics, Humboldt University, Berlin, Germany}

Within the Physics-Beyond-Collider initiative [1] and in the context of the European Strategy
of Particle Physics Update, a beam-dump facility (BDF) is being discussed at CERN’s SPS
[2]. The first experiment proposed at the BDF is the Search for Hidden Particles (SHiP)
experiment, see e.g.~[3]. ``Hidden” particles (such as heavy neutral leptons, dark photons/scalars, axion-like particles, and LDM particles) are predicted in many SM extensions with masses well below the electroweak scale and with
very small couplings to SM particles, which is why they might have escaped detection so far.

For SHiP, the high-intensity 400\,GeV SPS proton beam will be dumped in a heavy-metal
target. Hadrons emerging from the target are stopped in an absorber. Muons from hadron
decays are swept out by a magnet-based filter. SHiP consists of two parts: A 50 m-long
evacuated vessel followed by a spectro-/calorimeter allows one to reconstruct decays of long-lived
neutral particles produced in the SHiP target, which allows to search for hidden
particles through their decay with a very small expected background. Between the muon filter
and the decay vessel, the Scattering-and-Neutrino Detector (SND) is placed. One goal of the
SND is to study large statistics (anti-)tau-neutrino interactions as well as muon-neutrino
interactions to measure the strange-quark content of the proton and to search for charmed
pentaquark final states. Its special design allows SND to search in particular for LDM
particles that are potentially produced in the beam-dump target and are scattered off
electrons in the SND. An SND-like detector has been proposed to measure TeV-neutrinos of
all flavours produced in the very-forward directions of proton-proton collisions at the LHC.
This so-called SND@LHC experiment has been recently approved to be built and to take
data from 2022 onwards.

The presentation provided a general introduction to SHiP, discussed its prospects to search
for LDM particles below a few 100\,MeV and ended with the SND@LHC project.

\subsection{DM Searches at Jefferson Lab}

\speaker{Stepan Stepanyan, Thomas Jefferson Nation Accelerator Facility, USA}

The overwhelming evidence for DM in cosmological observations,
manifested by its gravitational interactions, has inspired a major experimental effort
to uncover its particle nature. The LHC, as well as direct and indirect detection
experiments, have significantly constrained one of the best-motivated weak-scale DM
models (WIMPs as DM candidates). In contrast, scenarios involving a light
hidden sector DM with masses in the MeV--GeV range has garnered a good
deal of attention. Models with hidden $U(1)$ gauge symmetry are particularly attractive as they can be tested experimentally. If these vector gauge bosons or dark/heavy
photons exist, they mix with ordinary photons through kinetic mixing, which induces
their weak coupling to electrons, $\epsilon e$. Since they couple to electrons, heavy photons
are radiated in electron scattering and can subsequently decay into $e^+e^-$ or to a pair
of LDM particles. Experiments at Jefferson Lab use these signatures to
search for heavy photons or LDM particles in the MeV to GeV mass
range.

In this talk, the experimental program and introduce facilities
at Jefferson Lab for DM searches were summarized.

\subsection{Discussion of Fixed-Target Experiments}

\speaker{These points were compiled by the scientific organizers. Any statements, opinions, or conclusions contained herein do not necessarily represent the view of the speakers or participants. They are not meant to be complete or representative, but highlight the lively interchange of ideas in the seminar.}

\begin{itemize}[leftmargin=*]

\item {\bf BDX-mini Experiment}
The BDX-mini experiment took data downstream of  the JLab experimental Hall-A beam dump for characterizing
the background conditions for the planned BDX experiment.
Two wells were drilled behind the beam dump one after the other with 3\,m spacing on the
beam axis.\\
Beam-off periods were used for background measurements, since it was not possible
to gate the experiment to the CEBAF beam structure: one bunch every 4\,ns makes it
hard to implement any time coincidence between signals in the BDX-mini calorimeter and the 
RF signal from the accelerator, given a time resolution of few ns. 
Beam-related backgrounds (e.g.\ neutrinos) could not be removed by time coincidence, since 
the time-of-flight for a GeV-energy neutrino and a GeV-energy LDM particle are basically 
the same for the $\sim 20$\,m separation between the beam dump and the detector.\\
Concerning beam-related neutron background, it was demonstrated that no high energy particles would reach the detector location consistently with  MC simulations  suggesting that
all high-energy neutrons would be absorbed by the  shielding. 
The effect of the low-energy neutron field on the detector (crystals, plastic scintillators, and photo-sensors) 
was experimentally checked with BDX-mini without observing critical issues.

\item {\bf NA62 Experiment}
NA62 has currently a trigger optimized for kaon decays and thus DM visible decays
are not favoured by the apparatus. For studying the response to DM paricles, biasing techniques
were implemented in Geant4. Other generators were considered (e.g.\ FLUKA, GENIE) but Geant4 was finally
chosen for the integration with existing tools within the experiment.\\
Currently, there is an intense on-going activity in NA62 for developing improvements for reaching the 
theoretical uncertainty for the ultra-rare $K\rightarrow \pi \nu \bar{\nu}$ decay.
The proposed KLEVER experiment will aim for the even more challenging $K^0\rightarrow \pi^0 \nu \bar{\nu}$ 
decay in future.

\item {\bf DarkMESA Experiment}
The DarkMESA veto system will be fully hermetic, thus also non-vertical cosmic rays will be detected
and vetoed. The efficiency of the veto detector is currently studied with a realistic simulation
of the cosmic ray spectrum (CRY).
The information from the veto system will be used off-line for fully exploiting the data,
since the expected rate will be low.\\
Concerning the detector simulation, it was suggested that
there is a way to improve the interaction of DM particles with atomic electrons in the detector 
which is implemented e.g.\ in GENIE and which takes care also of the inhomogeneities and multiple materials.\\
The DRIFT negative-ion time projection chamber can be an additional experimental technique for LDM detection applied at MESA. 

\item {\bf MAGIX Experiment}
The combinatorial background in the MAGIX experiment can be relevant at low energies for electron-positron coincidence and thus for visible dark photon decay searches.
Currently this background is not included in the simulation but there are plans to include it.\\
No veto systems dedicated to photons (or neutrals in general) are considered at the moment.
These issues were found to be relevant during the studies for the DarkLight experiment.
It should be straightforward to simulate the Bethe-Heitler background $e^-p\rightarrow e^-p\gamma$
and understand this contribution.\\
Currently the scattering angle resolution is limited and magnetic optics studies are underway. 
The $e^+e^-$ invariant mass resolution is driven by the opening angle, i.e.\ the in-plane angle is more important to be measured than the out-of-plane angle. 

\item {\bf SHiP Experiment}
SHiP will employ emulsion detectors, which were already successfully used in the past
(e.g.\ by the OPERA experiment). The plan is that emulsion detectors will be inspected in an automated 
way twice/year to keep the track density below $10^5/$cm$^2$.

\item {\bf Data Storage for DS$/$DM Experiments}
There is still no central place where upper limits for dark photon$/$LDM searches are stored. 
For example, the direct detection community has such a place at
\url{https://supercdms.slac.stanford.edu/dark-matter-limit-plotter}.
Another solution is \url{https://www.hepdata.net}.
A data storage and repository for DS searches could be a 
topic for the Initiative for DM in Europe \& beyond, see \url{https://indico.cern.ch/event/1016060}. 
Another existing project for axion-like particles (ALPs) and dark photons is \url{https://github.com/cajohare/AxionLimits}.\\

\item {\bf LDMX Experiment}
Photo-nuclear reactions are a subtle and key background for LDMX. A data-driven approach with 
control regions will be used also to understand and check simulations which are not accurate right now for this. 
In addition, different generators (Geant4, MCNP, FLUKA, PHITS) will be compared.\\
Other kinds of calorimeter technologies were considered, but budget and size constraints led to a sampling calorimeter with LHC/CMS technology.\\
The tagging tracker+magnetic field is very efficient in filtering off-beam particles but this aspect has to be studied in more depth.\\
The sensitivity to axion-like particles coupled to photons with the calorimeter is not very competitive with respect to other experiments. Projections can be found at \eprint{1807.01730}.

\item {\bf PADME Experiment}
The PADME active target did not experience any degradation until now.
The collaboration is planning for future experiments with positron beams and thin targets.
A key issue for such experiments is to stretch the beam and realize particle-by-particle detection, an on-going effort at Frascati. JLab might have a high-energy positron beam up to 11\,GeV
and this would be great for extending the mass reach. The time scale for this project is $\sim$5 years.

\item {\bf MUSE Experiment}
MUSE plans to use a MHz beam structure (one particle/bunch). If this beam will be available,  PADME-like measurements might be possible with 
a small-angle calorimeter. 
The MUSE 1-MHz beam will have a high $e^+$ fraction at energies of $100 - 250$\,MeV. 
The Collaboration might need to consider a special trigger condition, i.e.\ triggering on the 
beam particle with upstream detectors, and vetoing against beam particles with the downstream detectors. 
A key requirement is that beam-induced backgrounds should be small. 

\end{itemize}

\section{Direct-detection Experiments}

\subsection{DEAP-3600 and the Global Argon DM Programme}

\speaker{Tina R. Pollmann for the DEAP-3600 Collaboration,
NIKHEF/University of Amsterdam, Science Park, Amsterdam, Netherlands}

The DEAP-3600 experiment is designed to look for elastic scattering of WIMP DM on argon, using a 1 tonne liquid argon fiducial volume in single-phase
configuration. DEAP-3600 finished its first, four-year science run last year,
demonstrating excellent stability and superb pulse-shape discrimination power
against electromagnetic backgrounds. The DEAP-3600 detector is currently being
upgraded for a second science run and R\&D work toward future detectors.

DEAP-3600 will be followed by the DarkSide-20k experiment that is currently being
constructed by the Global Argon DM Collaboration (GADMC). The ultimate
goal of GADMC is a $\cal O$(100 tonnes) detector with WIMP sensitivity reaching to the
neutrino-mist region in parameter space for WIMP masses above 10\,GeV.

\subsection{LDM Search with the CRESST-III
Experiment}

\speaker{Federica Petricca, Max-Planck-Institut für Physik, München, Germany}

CRESST (Cryogenic Rare Event Search with Superconducting Thermometers) is a
direct DM search experiment located at the Gran Sasso underground
Laboratory (LNGS, Italy). Scintillating CaWO$_4$ crystals, operated as cryogenic
calorimeters at millikelvin temperature, are used as target material for elastic DM--nucleus
scattering. The experiment, optimized for low-energy nuclear recoil detection,
reached an unprecedented threshold of 30\,eV [1] for nuclear recoil energies and it is
currently leading the field of LDM search for values below 1.6\,GeV$/c^2$. 

In this contribution, the current stage of the CRESST-III experiment, together
with the most recent DM results were presented. The perspective for the
next phase of the experiment was also discussed.

\subsection{LDM Searches with Cryogenic Silicon
Detectors}

\speaker{Belina von Krosigk, Institut für Experimentalphysik, Universität Hamburg, Hamburg, Germany}

A new era has begun for direct DM searches using cryogenic silicon
detectors with reaching single electron-hole pair sensitivity. The corresponding ultralow
threshold and high resolution allow to search for LDM candidates of
masses as low as 500\,keV. This reach is possible through the DM electron
scattering detection channel, a channel that was inaccessible for these detectors
before. 

An overview on respective state-of-the-art direct detection
experiments was given and the current searches for LDM with electron
recoil signatures in the detectors was reviewed. The talk was concluded with an outlook on where the next
few years are expected to take us in this quest.

\subsection{LDM Searches with XENON1T}

\speaker{Michelle Galloway for the XENON Collaboration, University of Zurich, Zurich, Switzerland}

The XENON1T experiment was designed primarily to detect interactions of GeV-scale WIMP DM from its recoil off of a
xenon nucleus. To reach high sensitivity for WIMP searches, an ultralow background
for electronic recoils was achieved, thus enabling searches for LDM
candidates, such as dark photons and axion-like particles [1,2]. 

In this talk, the most recent results from these searches with XENON1T were presented and the reach
of xenon-based detectors to probe the LDM parameter space was exemplified.

\subsection{Discussion of Direct Detection Experiments}

\begin{itemize}[leftmargin=*]

\item {\bf DEAP-3600}
Argon emits VUV light at 128\,nm and wavelength shifters have to be used.
DEAP employs TPB to shift the VUV light to 420\,nm. The DEAP collaboration
has characterized the TPB behavior and pulse shape within the experiment. Other
wavelength shifters are under study (e.g.\ PEN) for DEAP and future argon-based
experiments like DarkSide-20k.

\item {\bf CRESST-III}
The experiment employs a Gatti-Manfredi (or ``matched'') filter for optimal signal-to-noise
enhancement. 

\item {\bf SuperCDMS} 
Modern CCD-based technologies were discussed. A technology like skipper-CCDs,
which is currently successfully employed in LDM searches, e.g.\ in the SENSEI experiment, would be difficult
to use in other fields, such as astrophysics, where large area devices are 
typically required. 
Large skipper-CCDs are expensive to build and need to be cooled to strongly suppress leakage current for an ultralow threshold.\\
Mechanical noise is a relevant contribution to low-threshold experiments.
Vibrations especially contribute in phonon-based measurements and counteractions (like damping systems or seismic platforms) are being taken. EM noise (and also vibrations to some extend) can further be suppressed by pulse-shape discrimination.\\
Recent excesses in very-low threshold experiments were discussed, where a careful
analysis is still on-going.

\item {\bf XENON-1T} The calculation for xenon is simple and does not require the use of QED. Instead, Hartree-Fock methods are used to compute the electronic wave functions.
Materials beyond Si are in the calculations such as Ge, NaI, CsI, GaAs.\\
Directional detection capability is possible for the NEXT detector, however, this is a high-pressure gas TPC that looks for $0\nu\beta\beta$, i.e.\ $\sim 2$\,MeV signals. It could be that this will be looked at also for DM searches. 

\end{itemize}

\section{Collider Experiments}

\subsection{LDM Searches at (Super) B-Factories} 
\speaker{Torben Ferber, Deutsches Elektronen-Synchrotron (DESY), Hamburg}

The B-Factories Belle and BaBar provided an excellent environment to
search for light mediators in the GeV mass range at e$^+$e$^-$ colliders. First, an overview was given of the searches at Belle and BaBar. In the second
part the sensitivity and first results of the Super B-Factory
Belle~II that started data taking in 2019 were described. With dedicated triggers for light
mediator searches, Belle~II offers unique opportunities to search for
axion-like particles, light $Z^\prime$, invisible final states, and extended DM models with long-lived signatures.

\subsection{DM Searches at the Large Hadron Collider and Synergies with other Experiments}

\speaker{Caterina Doglioni, Fysikum, Lund University, Lund, Sweden}

One overarching objective of science is to further our understanding of the universe,
from its early stages to its current state and future evolution. This depends on gaining
insight on the universe’s most macroscopic components, for example galaxies and
stars, as well as describing its smallest components, namely elementary particles
and nuclei and their interactions. The apparent excess of DM in the
universe remains one of the outstanding questions in science. If DM is a
particle, then it can be produced and sought at the Large Hadron Collider,
complementing searches in other experiments. 

This talk focused on the searches for
DM by experiments at the Large Hadron Collider [1], with a special highlight
on the searches that allow to probe DM hypotheses that are complementary
to other experiments.

It is clear that solving the DM puzzle requires combined expertise from the
fields of particle physics, astroparticle physics and nuclear physics.
Pursuing common scientific drivers such as DM also requires mastering
challenges related to instrumentation (e.g. beams and detectors), data acquisition,
selection and analysis, as well as making data and results available to the broader
science communities. This contribution also presented the work that various
communities and experiments are doing in this direction, and the ongoing initiatives
aiming to exploit synergies across different communities.

\subsection{DS Searches at BESIII}

\speaker{Dayong Wang,
School of Physics and
State Key Laboratory of Nuclear Physics and Technology, Peking University, Beijing, China}

Low energy, high luminosity $e^+e^-$ colliders are believed to be good places to search
for some exotic particles predicted in new physics models with DS
phenomenology. BESIII as the only currently running tau-charm factory has great
potential to probe these particles and models, with the largest samples of directly
produced charmonia. 

In this talk, some of such searches and related results were reported,
including dark photon searches using both the initial state radiation and $J/\psi$ decays
in association with a pseudoscalar meson ($\eta, \eta^\prime$), the searches for a lepto-phobic
dark photon, light Higgs etc, and more generally the study of processes with invisible
signatures, such as invisible decays of vector meson ($\omega, \phi$), pseudoscalar
mesons ($\eta$, $\eta^\prime$), charm mesons, hyperons ($\Lambda$), $J/\psi \rightarrow \gamma +$ invisible etc.

\subsection{Discussion of Collider Experiments}

\speaker{These points were compiled by the scientific organizers. Any statements, opinions, or conclusions contained herein do not necessarily represent the view of the speakers or participants. They are not meant to be complete or representative, but highlight the lively interchange of ideas in the seminar.}

\begin{itemize}[leftmargin=*]

\item One of the advantages of Belle~II with respect to e.g.\ BaBar is the staggering
of the crystals of the EM calorimeter which are not aligned towards the
interaction point and thus do not have ``blind'' lines of sight. This
brings roughly a factor 5 improvement in the limits for dark photon searches.\\
At Belle~II, several displaced vertex analyses are on-going.

\item Often trigger conditions and vetoing strategies are not fully reported in the
literature which does not help in comparing exclusion limits from different experiments.
In future publications efficiency maps, trigger conditions, and event selections must be included.

\item BESIII will also look for invisible dark photon decays. With this aim
a single-photon trigger was deployed in 2021, although understanding efficiencies will take time.\\
Searches for very light dark photons are difficult because of the large pion background. At the moment there is a focus on phase space regions that are competitive with BaBar.

\end{itemize}

\section{Other Experiments}

\subsection{X17 and the Search for New Physics in Nuclear Transitions}

\speaker{Viktor Zacek, Université de Montréal, Montréal, Canada}

Nuclear transitions provide a means to probe light, weakly-coupled new physics and
portals into the DS. Particularly promising are those transitions that can be
accessed through excited nuclear states that are resonantly produced, providing a highstatistics
laboratory to search for MeV-scale new physics. 

In this talk the so-called X-17 anomaly was discussed, which is a 7-$\sigma$ discrepancy reported by the ATOMKI group in
the observation of the decays of excited $^8$Be and $^4$He nuclei to their ground states via
internal $e^+e^-$ pair creation. The anomaly can be explained by the emission of a neutral
boson with a mass of about 17\,MeV$/c^2$. The ATOMKI results and their interpretations were
discussed, as well as follow-up experiments, among which an ongoing project at the
Montreal tandem accelerator facility.

\subsection{DarkLight@ARIEL: A Search for New Physics with
Invariant Mass between 10 and 20 MeV}

\speaker{Jan C. Bernauer for the DarkLight Collaboration, Center for Frontiers in Nuclear Science, Stony Brook University, Stony Brook, NY, USA, and RIKEN BNL Research Center, BNL, Upton, NY, USA}

Motivated by the recent anomalies found in $^8$Be and $^4$He, as well as muon $g-2$, the
DarkLight@ARIEL experiment aims to search for a DS particle with preferential
leptonic coupling. To this end, the experiment will measure the invariant mass spectrum of
e$^+$e$^-$ pairs produced in electron scattering off a thin tantalum target. Optimized for the search
around the predicted mass, the experiment will make use of the high intensity electron beam
of ARIEL.

In the talk, the motivation and history of the DarkLight experiment,
as well as the current design and status were presented.

\subsection{Status of the TREK/E36 Experiment at J-PARC}

\speaker{Michael Kohl, 
Physics Department, Hampton University, Hampton, USA,
and Thomas Jefferson National Accelerator Facility, Newport News, USA}

Experiment TREK/E36 has collected stopped-kaon decay data at the J-PARC K1.1BR
beamline for a precision measurement of the ratio of decay widths $BR(K^+ \rightarrow e^+\nu)$ and
$BR(K^+ \rightarrow \mu^+ \nu)$, respectively, to test lepton universality, and to search for rare decay modes
producing light neutral bosons, which may serve as explanations for particle anomalies
and LDM. 

An overview of the experiment and analysis status was presented.

\subsection{Discussion of Other Experiments}

\speaker{These points were compiled by the scientific organizers. Any statements, opinions, or conclusions contained herein do not necessarily represent the view of the speakers or participants. They are not meant to be complete or representative, but highlight the lively interchange of ideas in the seminar.}

\begin{itemize}[leftmargin=*]

\item The X17 experiment at the University of Montreal will be able to search for
the proposed particle in different nuclei. Currently, the read-out electronics
for the DAPHNE cylindrical drift chamber is tested. The cylindrical symmetry
of the setup will allow for a smoother acceptance distribution with respect to
the original ATOMKI experiment.

\item DarkLight abandoned the cylindrical symmetry design initially foreseen at JLab
because lepton tracks of interest are generally low in momentum and would curl up in the solenoidal field, eventually not exceeding a certain radius. In that case they would not be available for a trigger.

\end{itemize}

\section{Posters}

\speaker{(in alphabetical order)}

\subsection{The MAGIX Jet-Target System}
\speaker{Stephan Aulenbacher for the MAGIX Collaboration, Institute for Nuclear Physics, Johannes Gutenberg University Mainz, Germany}

For the planned LDM experiments at MAGIX a high precision is one of the
most crucial goals. Since the passage of electrons through matter
is introducing uncertainties on the angle and the energy of the final as well as on the
initial electrons, one has to keep the material budged low. Therefore, we implemented of a so-called jet target, developed by the University of Münster.
This target realizes a completely windowless scattering of electrons on a gas target,
by shooting a gas jet through the vacuum chamber.
This poster showed an overview on the entire target system, from the gas source to the
exhaust of the gas. It gave a short explanation on each subsystem of the MAGIX
target section.

\subsection{\texorpdfstring{\boldmath An Experimental Setup for Detection of $e^+e^-$ Pairs in the Decay of $^8$Be$^*$}{An Experimental Setup for Detection of e+e- Pairs in the Decay of 8-Be*}}

\speaker{Riccardo Bolzonella, 
Istituto Nazionale di Fisica Nucleare (INFN), Legnaro and Università di Padova, Dipartimento di Fisica e Astronomia, Padova, Italy
}

Recent papers of A.J. Krasznahorkay and collaborators proposed the existence of a
light neutral boson, named X17, for the interpretation of anomalies in the correlation
angle distribution between $e^-$ and $e^+$ emitted in the decay of excited states in $^8$Be and
$^4$He.

A new experimental setup is being developed at the National Laboratories of Legnaro
to provide an independent measurement of the effect.

This contribution focused on the design of a dedicated setup, describing the
detector layout, the simulation work done for its optimization and the experimental
characterization of the first prototypes.
The current layout is constituted by $\Delta$E-E organic scintillator telescopes, gathered in
groups of four, whose dimensions have been optimized, with the goal of improving
the angular resolution obtainable. The telescopes are read out by Silicon
PhotoMultipliers (SiPMs), that allow to keep small dimensions to fit the detectors in a
scattering chamber and would also be compatible with a future use within a magnetic
field.

The first layer of the telescopes is used to gather information on the particles'
positions: it is composed by 2 layers of orthogonal bars, which allow to measure the
two coordinates of the entry position and the $\Delta$E deposited energy. Also the bars are
read with an array of SiPM, for which an innovative readout scheme is proposed.

As a first step, a complete simulation of the setup was discussed. It has been
performed using the GEANT4 package to optimize the geometry and estimate the
detection efficiency.
Moreover, a preliminary characterization of the detector prototypes was discussed.
This characterization allows to estimate the expected resolution on the energy
measured by a detector and the resolution on the reconstructed invariant mass.
Eventually, this work is the preparation of a test beam that will be performed to
observe the pairs produced in the $^{19}\mathrm{F}(p,\alpha e^+e^-)^{16}\mathrm{O}$ reaction, to characterize in-beam
the prototype detectors together with the acquisition and analysis systems.

\subsection{
Studies on the Performance of the MAGIX Jet Target
}

\speaker{\underline{Philipp Brand}, Sophia Vestrick, and Alfons Khoukaz,
Institut für Kernphysik, WWU Münster, Germany}

The MAGIX experiment is a versatile experiment which allows, e.g., to search for
dark photons, and to measure the proton radius and astrophysical S-factor. To allow
for this challenging experimental program the powerful, future energy recovery linac
MESA will be used in combination with a jet target, that allows target thicknesses of
more than 10$^{18}$\,atoms$/$cm$^2$. This target was constructed and built up at the University
of Münster and is already in routine operation at MAMI in Mainz.

Extensive studies have been performed with the MAGIX target at the existing MAMI
facility to analyze and improve the jet target performance. This includes a simulation-based
optimization of the nozzle shape to reduce the divergence of the resulting
supersonic gas jet, which then also has a positive influence on the vacuum
conditions within the scattering chamber. The results of this optimization process were presented and discussed by comparing jet profiles and vacuum conditions for
different nozzle designs.

\subsection{Illuminating the Dark Photon with Darklight}

\speaker{Ethan Cline, 
Department of Physics and Astronomy, Stony Brook University, Stony Brook, USA}

The search for a dark photon holds considerable interest in the physics community
as such a force carrier would begin to illuminate the DS. Many experiments
have searched for such a particle, but so far it has proven elusive. In recent years the
concept of a low mass dark photon has gained popularity in the physics community.
Of particular recent interest is the $^8$Be anomaly, which could be explained by a 17\,MeV mass dark photon. The proposed Darklight experiment would search for this
potential low mass force carrier at TRIUMF in the $10-20$\,MeV $e^+e^-$ invariant mass
range. This poster focused on the experimental design and physics case of the
Darklight experiment.

\subsection{Search for Light Neutral Bosons in the TREK/E36 Experiment}

\speaker{Dongwi H. Dongwi (Bishoy)}

\subsection{The Silicon Strip Detector Setup for MAGIX}

\speaker{Jennifer Geimer for the MAGIX Collaboration, 
Institute for Nuclear Physics, Johannes Gutenberg University Mainz, Germany
}

The MAGIX (Mainz Gas Internal Target Experiment) experiment will take place at the
energy recovering superconducting accelerator MESA in Mainz.
At MAGIX, high-precision electron scattering experiments will be performed covering
a wide experimental program like investigations of hadron physics, reactions of astrophysical
relevance as well as DS searches.

The experimental setup is currently under development and provides a windowless gas
jet target and two identical high-resolution magnetic spectrometers including a GEM-based
time projection chamber.

Additionally, a silicon strip detector is planned to detect recoil particles inside the scattering
chamber. Its main requirements are suited to the simulation of the S-factor
determination of the nucleosynthesis reaction of carbon and alpha which defines the
lower limit of the energy sensitivity to 0.3\,MeV.

Other reactions like the invisible decay of the dark photon $p(e,e'p)\chi\overline{\chi}$ increase the
needed energy range of the recoil detector to several MeVs.
Therefore the silicon detector will be extended by an additional plastic scintillator layer.

The current state of the silicon strip detector development and its underlying working
concept were presented on this poster.

\subsection{Shining Light into the Dark (Jets) with ATLAS}

\speaker{\underline{Jannik Geisen} and Caterina Doglioni, Department of Physics, Lund, Sweden}

The Large Hadron Collider (LHC) at CERN, Switzerland, reached the conclusion of its
second data taking period from 2015 to 2018, and with that produced the largest
proton--proton collision particle physics dataset to date. These data are being
analysed by the ATLAS experiment with ever increasing precision and even more
sophisticated strategies. In particular, discovering the nature of DM in high
energy proton--proton collisions is one of the experiment’s major goals.

A new approach to search for DM in the full Run-2 dataset with
the ATLAS experiment was presented. Similar to the SM, a DS could
exist in the Universe, containing new particles as well as new interactions such as a
dark version of QCD. Dark QCD includes dark quarks
which could be produced at the LHC. These dark quarks undergo a dark showering
and hadronisation process inside the ATLAS detector producing a large number of
light dark and/or SM hadrons bundled together into jets. The search that was presented
exploits the internal structure of such dark jets. Furthermore, since the DS
could be manifested in different ways~[1] resulting in different detector signatures, the phenomenological studies were highlighted which depict these features and offer
optimization strategies.

\subsection{Simulation of Exclusion Limits for the Invisible Decay of Dark Photons at MAGIX}

\speaker{Pepe Gülker for the MAGIX Collaboration, Institute for Nuclear Physics, Johannes Gutenberg University Mainz, Germany}

With the planned MAGIX experiment in Mainz a versatile apparatus will be available to
search for radiatively produced dark photons with a mass below 100\,MeV$/c^2$ down to a
coupling constant $\epsilon$ smaller than $10^{-5}$, a region that has yet to be proped intensively.

The missing mass spectra will be recorded by the unique electron scattering setup consisting
of two rotatable high resolution magnetic spectrometers, and a set of dedicated
recoil detectors surrounding a central gas jet target. The very intense electron beam
of up to 10\,mA will be provided by the energy recovering superconducting accelerator
MESA.

This contribution focused on the simulation of the invisible decay channel and showed
the path from the generators to the extraction of exclusion limits in detail. It also provided
the theoretical framework and motivated the requirements for hardware related
contributions in the scope of MAGIX presented during this seminar.

\subsection{Light. Dark. Resonant: Sub-GeV Thermal DM}

\speaker{\underline{Saniya Heeba}, Elias Bernreuther, and Felix Kahlhoefer, Institute for Theoretical Particle Physics and Cosmology, RWTH Aachen University, Aachen, Germany}

The particle-physics description of DM remains an open question in high
energy physics and cosmology. With increasing sensitivities of various experiments
and an absence of a clear DM signal, alternatives to the conventionally
studied WIMP paradigm have gained traction. One class of such alternatives looks at
light (sub-GeV) DM. These models necessarily require some alteration to the
DM production history to ensure compatibility with cosmological bounds such as
those coming from the CMB. One can do this by considering models in which dark
matter annihilation is resonantly enhanced during freeze-out but suppressed at CMB
times. Interestingly, the viable parameter space in such models is testable by direct
detection and beam dump experiments in complementary ways making them a good
target for current and future DM searches.

This poster illustrated the theoretical intricacies of such models and highlighted the
different approaches one can take to probe them experimentally.

\subsection{The DarkMESA Veto Detector}

\speaker{\underline{Matteo Lau\ss}, Patrick Achenbach, Mirco Christmann, Luca Doria, Manuel Mauch and Sebastian Stengel for the MAGIX-Collaboration, Institute for Nuclear Physics, 
Helmholtz Institute Mainz and PRISMA+ Cluster of Excellence, Johannes Gutenberg University Mainz, Germany}

At the Institute for Nuclear Physics in Mainz the new electron accelerator MESA will go
into operation within the next years. In the extracted beam operation (150\,MeV, 150\,$\mu$A)
the P2 experiment will measure the weak mixing angle in electron-proton scattering in
10\,000 hours operation time. Therefore, the high-power beam dump of this experiment
is ideally suited for a parasitic DS experiment --- DarkMESA [1,2].

The experiment is designed for the detection of LDM which in
the simplest model couples to a massive vector particle, the dark photon 
$\gamma'$. It can
potentially be produced in the P2 beam dump by a process analogous to photon
bremsstrahlung and can then decay in DM particle pairs $\chi\overline{\chi}$. A fraction
of them scatter off electrons or nuclei in the DarkMESA calorimeter [3].

The better the suppression of background radiation (e.g.\ cosmic muons) the better the
sensitivity of the experiment. Therefore, a highly efficient veto detector surrounding the
calorimeter hermetically is essential to probe the target parameter space of DarkMESA
successfully. The veto detector will consist of two layers of plastic scintillation counters
separated by a lead layer. To test the characteristics of such a detector a prototype
is currently under construction using 2\,cm thick plastic scintillators of type EJ-200 and
a matrix of $5 \times 5$ lead fluoride crystal bars as the calorimeter. The efficiency of the
veto system at shielding the calorimeter from cosmic background radiation has been
simulated using the Geant4 software. Results and practical implications were discussed.
Furthermore, the option of using Gadolinium loaded scintillators to increase
the detection efficiency of slow neutrons will be examined. The production of beam related
neutrons is currently being studied within a FLUKA simulation showing no thermal
neutrons reaching the DarkMESA calorimeter.

\subsection{MAGIX Slow Control}

\speaker{Stefan Lunkenheimer for the MAGIX Collaboration, Institute for Nuclear Physics, Johannes Gutenberg University Mainz, Germany}

MAGIX (Mainz Gas Injection Target Experiment) is a versatile fixed-target experiment
and will be built at the new electron accelerator MESA (Mainz Energy-Recovering Superconducting
Accelerator) in Mainz. The accelerator will provide a polarized and unpolarized
electron beam with a current up to 1\,mA and a beam energy up to 105\,MeV.
Using its internal gas jet target, MAGIX will reach a luminosity of ${\cal O} (10^{34}$\,cm$^{-2}$s$^{-1}$). In
a rich physical program MAGIX allows to study processes with very low cross sections
at small momentum transfer.

The existence of a dark photon that acts as an exchange particle for DM is a
well-motivated extension of the SM of particle physics. Through the mixing
of the dark photon with the ordinary photon, the dark photon can couple very weakly to
charged baryonic particles. One possible way to identify the dark photon is to perform
an electron scattering experiment to determine the missing mass of the invisible decay
reaction: $p(e,e'p)\chi\overline{\chi}$. MAGIX can use the high-resolution spectrometer in combination
with the silicon-strip detector array to take part in the search for the radiatively produced
dark photon. Together with the thin gas jet target the kinetic parameters and thus the
missing mass can be determined with high precision. For such an high-precision
measurement, it is necessary to control all parameters with high accuracy. MAGIX
therefore needs a comprehensive slow control system for the entire experiment.

On the poster the requirements for such a slow control system were presented as well
as the current developments. MAGIX will use an EPICS-based slow control system
that is already applied for the existing MAGIX components. EPICS is a decentralized
system with which all parts for the experiment can simply be separated and combined.
MESA and DarkMESA also use EPICS, which means that all parameters can easily
be shared between the experiments and the accelerator. In the current test stands the
MAGIX slow control system controls a total of around 1500 parameters.

\subsection{Dark Photon Production Via Positron Annihilation In Electron-Beam Thick-Target Experiments}

\speaker{Luca Marsicano, INFN Genova, Italy}

In a popular class of models, DM is composed of particles with mass in the
MeV--GeV range, interacting with the SM via a new force, mediated by a
massive vector boson, the dark photon or $A'$. High energy positron annihilation is a
viable mechanism to produce dark photons. This reaction plays a significant role in
beam-dump experiments using multi-GeV electron-beams on thick targets by
enhancing the sensitivity to $A'$ production. The positron-rich environment generated
by the electromagnetic shower initiated by the electron beam in the target allows to
produce a significant number of $A'$s via non-resonant ($e^+e^-\rightarrow \gamma A'$) and resonant
annihilation ($e^+e^- \rightarrow A'$) on atomic electrons. For both visible and invisible $A'$ decays,
the contribution of resonant annihilation results in a larger sensitivity with respect to
limits derived by the commonly used $A'$-strahlung in certain kinematic regions. The
sensitivity enhancement due to this process has been evaluated for different past
and proposed electron-beam thick-target experiments: E137, BDX, NA64 and LDMX.

This contribution presented the results of this study, with a detailed description of
the procedure adopted for the calculation.

\subsection{Readout System for DarkMESA Veto Detector}

\speaker{\underline{Manuel Mauch}, Patrick Achenbach, Mirco Christmann, Luca Doria, Matteo Lau\ss, and Sebastian Stengel for the MAGIX-Collaboration, Institute for Nuclear Physics, 
Helmholtz Institute Mainz and PRISMA+ Cluster of Excellence, Johannes Gutenberg University Mainz, Germany}

At the Institute for Nuclear Physics in Mainz a new beam dump experiment, called
DarkMESA, will go into operation during the next years. This experiment is designed
for the search for LDM particles with a crystal-based calorimeter. The detector
will leverage on over 10\,000 hours of available measuring time at the new MESA
electron accelerator [1,2].

A key issue will be the rejection of background events like cosmics, environmental radioactivity,
and beam particles. To this end, a veto detector system is being developed,
consisting of scintillator plates coupled to silicon photomultipliers (SiPMs).

The readout electronics currently developed is tested with the help of a prototype laboratory
setup which consists on a calorimeter comprised of a $5 \times 5$ crystal matrix, completely
enveloped in scintillator plates. The scintillators are read out with specifically
designed ``carrier boards'' on which are installed SiPMs and corresponding preamplifiers.
The signals of several carrier boards will be transferred to a second ``collector''
board.

It will be possible to further process the signals on an external FPGA board while the
analog signals will be digitized on a sampling ADC. First measurements with the carrier
and collector boards were presented, and concepts for the FPGA boards and the
ADC were discussed.

\subsection{Audible Axions}

\speaker{Wolfram Ratzinger, PRISMA+ Cluster of Excellence and Mainz Institute for Theoretical Physics,
Johannes Gutenberg University Mainz, Germany}

Conventional approaches to probing axions and axion-like particles (ALPs) typically rely on a
coupling to photons. However, if this coupling is extremely weak, ALPs become invisible and are
effectively decoupled from the Standard Model. We show that such invisible axions, which are viable
candidates for DM, can produce a stochastic gravitational wave background in the early
universe. This signal is generated in models where the invisible axion couples to a dark gauge boson
that experiences a tachyonic instability when the axion begins to oscillate. Quantum 
fluctuations
amplified by the exponentially growing gauge boson modes source chiral gravitational waves. Lattice calculations of the resulting GW signal were presented and its detectability as well as
the possibility of explaining the recent NANOGrav result were highlighted. Finally, it was shown that this mechanism
can produce GWs in a wide range of axion scenarios, considering the relaxion and an kinetically
misaligned axion.

\subsection{Charming ALPs}

\speaker{\underline{Christiane Scherb}, Adrian Carmona, and Pedro Schwaller, CAFPE and Departamento de F\'isica Teórica y del Cosmos,
Universidad de Granada, Spain
and PRISMA+ Cluster of Excellence \& Mainz Institute for Theoretical Physics, Johannes Gutenberg University Mainz, Germany}

Axion-like particles (ALPs) are ubiquitous in models of new physics explaining some
of the most pressing puzzles of the SM. However, until relatively
recently, little attention has been paid to its interplay with flavour. In this work, we
study in detail the phenomenology of ALPs that exclusively interact with up-type
quarks at the tree-level, which arise in some well-motivated ultra-violet completions
such as QCD-like DS or Froggatt-Nielsen type models of flavour. Our study
is performed in the low-energy effective theory to highlight the key features of these
scenarios in a model independent way. We derive all the existing constraints on
these models and demonstrate how upcoming experiments at fixed-target facilities
and the LHC can probe regions of the parameter space which are currently not
excluded by cosmological and astrophysical bounds. We also emphasize how a
future measurement of the currently unavailable meson decay $D \rightarrow \pi +$ invisible could
complement these upcoming searches. For small masses the charming ALP is a DM candidate.

\subsection{Position Reconstruction in DEAP-3600 Experiment Using Neural Networks}

\speaker{\underline{Goran Stanić} and Luca Doria, Institute for Nuclear Physics, Johannes Gutenberg University Mainz, Germany}

The DEAP-3600 is a direct-detection DM experiment, located in the SNOLAB
facility in Sudbury, Canada. With its spherical acrylic vessel filled with 3.3\,tonnes of
single-phase liquid argon, it aims at detecting spin-independent WIMP--nucleon scattering.
Argon scintillation light (wavelength-shifted by TPB coating) is detected by 255 PMTs
arranged around the vessel.
Position reconstruction in DEAP-3600 is of utmost importance for background
rejection and fiducialization [1].

Existing algorithms are based on the charge or time pattern of the PMTs [2].
The goal of this work is to train and test two neural network architectures --- the
Feedforward and the Convolutional Neural Network --- in order to achieve optimal
event position reconstruction and try to improve on existing algorithms.

The results obtained point to a precision of less than 50\,mm in position reconstruction
for both neural network architectures. The potential for identifying events coming
from the detector’s ``neck'' is also investigated.

\subsection{The MAGIX Trigger Veto System}

\speaker{Sebastian Stengel, Institute for Nuclear Physics, Johannes Gutenberg-University Mainz, Germany}

The MAGIX setup will be used for dark photon searches using the visible as well as
the invisible decay channel. The MAGIX trigger veto system will enable the fast
timing characteristics needed for investigating the visible dark photon decay channel
$A' \rightarrow e^+e^-$. It will further be used for energy-loss measurements and will provide the
basic hit and position information for the triggered readout of the MAGIX time
projection chamber.

The MAGIX trigger veto system will consist of one segmented trigger layer of plastic
scintillator bars and a flexible veto system of additional scintillation detectors and
lead absorbers placed below the trigger layer.
The data readout will use the ultrafast preamplifier-discriminator NINO chip
developed for use in the ALICE detector followed by FPGAs programmed as TDCs.

\subsection{\texorpdfstring{Münster Jet-Target for Future Dark Photon\\ Searches at MAGIX}{Münster Jet-Target for Future Dark Photon Searches at MAGIX}}

\speaker{\underline{Sophia Vestrick}, Philipp Brand, and Alfons Khoukaz, Institut für Kernphysik, WWU Münster, Germany}

The MAGIX experiment at MESA using a quasi-internal gas-jet target initially aims for
high precision measurements of scattering between the MESA electron beam and
various gases from the Münster jet target at low momenta.

One research topic from high interest is the search for dark photons, which can be
produced radiatively in the electron-nucleus-scattering. Precise measurements of this
dark photons require a high resolution of the MAGIX spectrometers and a gas-jet
target with a thickness of more than $10^{18}$\,atoms$/$cm$^2$, allowing for luminosities of up to
$10^{35}/$(cm$^2$s).
The MAGIX gas-jet target was build and tested in the Münster laboratories and is
currently installed at the A1 Experiment at MAMI. First beam times using this jet
target have been performed and showed that stable jet beam conditions with a target
thickness of $> 10^{18}$\,atoms$/$cm$^2$ have been confirmed. This proofs the excellent
suitability of this jet target for high precision, rare event measurements at MAGIX.

\end{document}